\documentstyle[12pt,graphicx,pstricks,braket,amsmath,amssymb,enumitem,
                    epstopdf,pdfpages,cite,tabu,hyperref,subfigure]{article}
\numberwithin{equation}{section}

\newcommand{\hi}[1]{}

\newcommand{\FbR}{{\bar F}_R}
\newcommand{\Db}{{\bar D}}
\newcommand{\Sb}{{\bar S}}
\newcommand{\phib}{{\bar \phi}}
\newcommand{\FL}{{F}_L}
\newcommand{\FR}{{F}_R}
\newcommand{\oh}{\frac{1}{2}}
\begin{document}

\def\AEF{A.E. Faraggi}

\def\ibid#1#2#3{{\it ibid} {\bf #1}, (#2) #3}
\def\JHEP#1#2#3{{\it JHEP} {\textbf #1}, (#2) #3}
\def\JCAP#1#2#3{{\it JCAP} {\textbf #1}, (#2) #3}
\def\vol#1#2#3{{\bf {#1}} ({#2}) {#3}}
\def\NPB#1#2#3{{\it Nucl.\ Phys.}\/ {\bf B#1} (#2) #3}
\def\PLB#1#2#3{{\it Phys.\ Lett.}\/ {\bf B#1} (#2) #3}
\def\PRD#1#2#3{{\it Phys.\ Rev.}\/ {\bf D#1} (#2) #3}
\def\PRC#1#2#3{{\it Phys.\ Rev.}\/ {\bf C#1} (#2) #3}
\def\PRL#1#2#3{{\it Phys.\ Rev.\ Lett.}\/ {\bf #1} (#2) #3}
\def\PRT#1#2#3{{\it Phys.\ Rep.}\/ {\bf#1} (#2) #3}
\def\MODA#1#2#3{{\it Mod.\ Phys.\ Lett.}\/ {\bf A#1} (#2) #3}
\def\RMP#1#2#3{{\it Rev.\ Mod.\ Phys.}\/ {\bf #1} (#2) #3}
\def\IJMP#1#2#3{{\it Int.\ J.\ Mod.\ Phys.}\/ {\bf A#1} (#2) #3}
\def\IJTP#1#2#3{{\it Int.\ J.\ Theor.\ Phys.}\/ {\bf A#1} (#2) #3}
\def\nuvc#1#2#3{{\it Nuovo Cimento}\/ {\bf #1A} (#2) #3}
\def\RPP#1#2#3{{\it Rept.\ Prog.\ Phys.}\/ {\bf #1} (#2) #3}
\def\APJ#1#2#3{{\it Astrophys.\ J.}\/ {\bf #1} (#2) #3}
\def\APP#1#2#3{{\it Astropart.\ Phys.}\/ {\bf #1} (#2) #3}
\def\EJP#1#2#3{{\it Eur.\ Phys.\ Jour.}\/ {\bf C#1} (#2) #3}

\def\etal{{\it et al.\/}}
\def\notE6{{$SO(10)\times U(1)_{\zeta}\not\subset E_6$}}
\def\E6{{$SO(10)\times U(1)_{\zeta}\subset E_6$}}
\def\highgg{{$SU(3)_C\times SU(2)_L \times SU(2)_R \times U(1)_C \times U(1)_{\zeta}$}}
\def\highSO10{{$SU(3)_C\times SU(2)_L \times SU(2)_R \times U(1)_C$}}
\def\lowgg{{$SU(3)_C\times SU(2)_L \times U(1)_Y \times U(1)_{Z^\prime}$}}
\def\SMgg{{$SU(3)_C\times SU(2)_L \times U(1)_Y$}}
\def\Uzprime{{$U(1)_{Z^\prime}$}}
\def\Uzeta{{$U(1)_{\zeta}$}}

\def\lsim{\raise0.3ex\hbox{$\;<$\kern-0.75em\raise-1.1ex\hbox{$\sim\;$}}}
\def\gsim{\raise0.3ex\hbox{$\;>$\kern-0.75em\raise-1.1ex\hbox{$\sim\;$}}}

\newcommand{\cc}[2]{c{#1\atopwithdelims[]#2}}
\newcommand{\bev}{\begin{verbatim}}
\newcommand{\beq}{\begin{equation}}
\newcommand{\bea}{\begin{eqnarray}}
\newcommand{\eea}{\end{eqnarray}}

\newcommand{\beqa}{\begin{eqnarray}}
\newcommand{\beqn}{\begin{eqnarray}}
\newcommand{\eeqn}{\end{eqnarray}}
\newcommand{\eeqa}{\end{eqnarray}}
\newcommand{\eeq}{\end{equation}}
\newcommand{\beqt}{\begin{equation*}}
\newcommand{\eeqt}{\end{equation*}}
\newcommand{\Eev}{\end{verbatim}}
\newcommand{\bec}{\begin{center}}
\newcommand{\eec}{\end{center}}
\newcommand{\bes}{\begin{split}}
\newcommand{\ees}{\end{split}}
\newcommand{\nn}{\nonumber}

\def\ie{{\it i.e.~}}
\def\eg{{\it e.g.~}}
\def\half{{\textstyle{1\over 2}}}
\def\nicefrac#1#2{\hbox{${#1\over #2}$}}
\def\third{{\textstyle {1\over3}}}
\def\quarter{{\textstyle {1\over4}}}
\def\m{{\tt -}}
\def\mass{M_{l^+ l^-}}
\def\p{{\tt +}}

\def\slash#1{#1\hskip-6pt/\hskip6pt}
\def\slk{\slash{k}}
\def\GeV{\,{\rm GeV}}
\def\TeV{\,{\rm TeV}}
\def\y{\,{\rm y}}

\def\l{\langle}
\def\r{\rangle}
\def\LRS{LRS  }

\begin{titlepage}
\samepage{
\setcounter{page}{1}
\rightline{LTH--1170}
\vspace{1.5cm}

\begin{center}
 {\Large \bf 
Sterile Neutrinos in String Derived Models
}
\end{center}

\begin{center}

{\large
Alon E. Faraggi\footnote{email address: alon.faraggi@liv.ac.uk}
}\\
\vspace{1cm}
{\it  Department of Mathematical Sciences,\\
             University of Liverpool,
         Liverpool L69 7ZL, UK\\}

\end{center}

\begin{abstract}

The MiniBooNE collaboration recently reported further evidence for the 
existence of sterile neutrinos, implying substantial mixing with the 
left--handed active neutrinos and at a comparable mass scale. 
I argue that while sterile neutrinos may arise naturally in large volume
string models, they prove more of a challenge in heterotic--string
models that replicate the Grand Unified Theory structure of the 
Standard Model matter states.
Sterile neutrinos in heterotic--string models 
may imply the existence of an additional Abelian gauge symmetry 
of order 10--100TeV.

\end{abstract}
\smallskip}
\end{titlepage}

\section{Introduction}

The MiniBooNE collaboration recently reported further evidence for the 
existence of sterile neutrinos \cite{miniboone}.
The MiniBooNE data strengthens the results of the LSND collaboration
that were obtained more than two decades ago \cite{lsnd}. 
If substantiated by further experimental data, these results will have
profound implications on string phenomenology. While sterile neutrinos
may arise naturally in large volume scenarios, they present more of a challenge
in heterotic--string models that replicate the matter structure 
of Grand Unified Theories (GUTs). It should, however, be stressed that
the experimental and phenomenological analysis in this regard is far from
settled \cite{dentler}, and highlights the urgency for further experimental 
data to resolve the question. 

Possible existence of sterile neutrinos has been studied
in the Beyond the Standard Model field theory constructions 
\cite{sterileft}. Neutrino masses in string 
inspired top--down scenarios is reviewed in \cite{topdown}. 
In this paper I examine the potential implications
of sterile neutrinos in string derived models. The discussion for the 
most part will be qualitative and more elaborate investigations are 
relegated for future work. 

Field theory extensions of the Standard Model often give rise 
to sterile neutrinos and by implications to neutrino masses and
mixing. The renormalisable Standard Model itself does not accommodate
neutrino masses, which are mandated by the observation of
neutrino oscillations \cite{neuosc}. However, as the sterile neutrinos
are neutral under the Standard Model gauge group, one can write for them
a large Majorana mass term, which produces the so--called seesaw mechanism
\cite{seesaw}. At low scales there remain three active neutrinos, which mostly 
consist of the Standard Model left--handed neutrinos, whereas the 
right--handed neutrinos mass is of the order of the seesaw scale. 
The seesaw mechanism naturally explains the suppression of the 
left--handed neutrino masses, compared to the mass scale of 
the charged Standard Model particles. 

This explanation of the light neutrino masses and oscillations also 
fits beautifully in $SO(10)$--GUT embedding of the Standard Model, 
in which each generation, augmented by a right--handed neutrino, 
fits in the chiral {\bf 16} representation of $SO(10)$. The structure
of the Standard Model matter charges then possess a robust mathematical
underpinning. The seesaw mass scale is tied to the GUT symmetry
breaking scale and therefore has a natural origin. Furthermore, GUT
mass relations between the neutral leptons and charged fermions
make the seesaw mechanism a necessity, rather than a nicety. 
Thus, within $SO(10)$--GUTs the neutrino spectrum generically consists 
solely of three active neutrinos. Extending the $SO(10)$ symmetry to $E_6$ 
may naturally give rise to sterile neutrinos, depending on the symmetry
breaking structures \cite{sterileft}. Furthermore, the existence of 
sterile neutrinos in this scheme may hinge on the existence of an 
Abelian symmetry, beyond the Standard Model, and that remains unbroken
down to low scales.

An alternative to the seesaw mechanism is the possibility that neutrinos 
obtain their mass via electroweak symmetry breaking. In this case 
sterile neutrinos naturally arise as right--handed neutrinos 
and mirrors the mass generation of the Standard Model charged sector. 
One then needs to explain the, at least, nine orders of magnitude suppression 
of the neutrino Yukawa coupling compared to the Yukawa couplings 
in the charged sector. In this case one also abandons the GUT scenario, 
which is motivated by Standard Model matter charges. 

In this paper, I focus for the most part on examination of sterile 
neutrinos in string GUT models. I argue that in the generic model
sterile neutrinos do not arise and discuss the conditions that 
may allow them to appear. String GUT models contain right--handed
neutrino states that arise from the {\textbf 16} spinorial representation
of $SO(10)$. While the $SO(10)$ gauge group is broken directly at the string 
level, remnants of the GUT symmetry give rise to mass relations between the 
up quark mass matrix and the neutrino Dirac mass matrix. Suppression
of the active neutrino masses therefore necessitates
employing the seesaw mechanism at a high scale, which gives the 
right--handed neutrinos mass of the order of the seesaw scale. 
Heuristically, we might associate the existence of light sterile neutrinos
with an unbroken Abelian gauge symmetry below the string scale. 
The caveat with string derived GUT constructions is that obtaining such 
extra $U(1)$ symmetries, that may remain viable below the string scale,
is not straightforward. The reason is that the typical $E_6$ $U(1)$ 
symmetries, that are amply discussed in the string inspired literature,
are generically anomalous in the string derived models \cite{cleaverau1}. 
A highly non--trivial construction that enables an additional $E_6$ $Z^\prime$
to remain unbroken below the string scale was presented in ref. 
\cite{frzprime}, and utilises self--duality under the spinor--vector
duality symmetry of ref. \cite{spinvecduality}.

\section{Sterile neutrinos in large volume scenarios}

In this section I briefly elaborate why sterile neutrinos may arise naturally
in large volume scenarios. Neutrino masses in these scenarios, in which 
the fundamental scale of quantum gravity may be as low as the TeV scale, 
were discussed in {\it e.g.} ref. \cite{lvneutrinos}. 
In this case the $SO(10)$ GUT embedding of the 
Standard Model states must be abandoned. As discussed above the reason is 
the $SO(10)$ mass relations that necessitate a high seesaw scale. 
In large volume scenarios the seesaw scale is low, indicating that the
origin of the right--handed neutrino states differs from that of the 
other Standard Model fields. This is achieved if the Standard Model 
particles are confined to a brane, whereas the right--handed neutrinos
can propagate in the bulk \cite{lvneutrinos}.
The neutrino mass terms arise from the higher dimensional kinetic terms 
that produce Dirac mass terms. 
The Yukawa couplings of the left-- and right--handed 
neutrinos are then suppressed by the volume of the extra dimensions.
Considering a five dimensional theory, $(x^\mu,y)$ with 
$\mu=0,\cdots,3$ and a compactified circle $y$ with radius 
$R$. The right--handed neutrino is bulk fermion state, while the
lepton and Higgs doublets are confined to the brane.
The bulk Dirac spinor is decomposed in the Weyl basis
$\Psi = \left( \nu_R , \bar{\nu^c}_R  \right)$
and takes the usual Fourier expansion
\begin{equation}
\nu_R^{(c)}(x,y) = \sum_n \frac{1}{\sqrt{2 \pi r}} \nu_{Rn}^{(c)}(x) e^{iny/r}
\end{equation}
The four dimensional model then contains
the usual tower of Kaluza--Klein states with Dirac
masses $n/r$ and a free action for the lepton
doublet, which is localized on the wall. 
The leading interaction term between the walls fields and the bulk
fermion is
\begin{equation}
S^{\rm{int}} = \int d^4 x \lambda l(x) h^*(x) \nu_R(x,y=0)
\label{intf}
\end{equation}
with $\lambda$ being a dimensionless parameter.
The Yukawa coupling $\lambda$ is rescaled like the 
graviton and dilaton coupling to all brane fields. 
The effective Yukawa on the four dimensional brane is given by
\begin{equation}
\lambda_{(4)}=\frac{\lambda}{\sqrt{r^nM_*^n}},
\end{equation}
which leads to very strong suppression of the Dirac mass even for $\lambda
\sim 1$:
\begin{equation}
m=\frac{v}{\sqrt{2}}\lambda_{(4)}=
\frac{\lambda v}{\sqrt{2}}\frac{M_*}
{M_{{\rm Pl}}}\simeq
\lambda\frac{M_*}{{\rm 1 TeV}} \cdot 5 \cdot 10^{-5}  {\rm eV}, 
\label{yuksin4d}
\end{equation}
where $v$ is the electroweak VEV. The brane left--handed neutrino 
couples to the tower of bulk Kaluza--Klein modes. The resulting 
mass matrix for every neutrino species is given by \cite{lvneutrinos}
\begin{equation}
{\bf M}=\left(
              \begin{array}{ccccccc}m&0&0&0&.&.&.\\
                                 m&1/r&0&0&.&.&.\\
                                 m&0&2/r&0&.&.&.\\
                                 m&0&0&3/r&.&.&.\\
                                 .&.&.&.  &.&.&.\\
                                 .&.&.&.  &.&.&.\\
                                 .&.&.&.  &.&.&.
              \end{array}
        \right)
\label{matrix}
\end{equation}
In the limit $m=0$, the mass matrix (\ref{matrix}) has one vanishing
eigenvalue and the usual ladder of KK masses. In this limit
the left-handed neutrino is decoupled from the
Kaluza--Klein tower of states. When $m$ is finite, 
the left-handed neutrino mixes with the other
states and the mixing angle $\theta_k$ between the left-handed 
neutrino and $k$-th Kaluza--Klein state is given by 
\begin{equation}
\theta_k\simeq \frac{mr}{|k|}
\end{equation}
The suppression of $\lambda_{(4)}$, similar to the suppression of the
gravitational couplings, results in the suppression of neutrino masses. 
However, for our purpose here it is noted that the tower of Kaluza--Klein
states acts naturally as sterile neutrinos and may lead to observable 
effects through their interactions with the left--handed neutrino masses
\cite{lvneutrinos}. String derived brane constructions that may be used 
toward realising the large volume scenarios were explored 
(see {\it e.g.} \cite{branemodels} and references therein). 

\section{Sterile neutrinos in string GUT constructions}

In this section I explore possible existence of sterile neutrinos in string 
GUT constructions, which are string models that retain the embedding 
of the Standard Model states in $SO(10)$ and $E_6$ representations.
While the GUT symmetries are broken in these models directly at the 
string level, the matter states are obtained from GUT representations
that are decomposed under the final unbroken GUT subgroup. Consequently, 
the embedding preserves the weak hypercharge $U(1)$ GUT charges, 
and the canonical $\sin^2\theta_W(M_{\rm GUT})=3/8$ normalisation. 
String GUT 
models may, in general, be obtained from perturbative and nonperturbative 
constructions.
The concrete class of models that are investigated here 
are perturbative heterotic string models in the free fermionic formulation
\cite{fff}. These string vacua correspond to $Z_2\times Z_2$ toroidal 
orbifold compactifications \cite{z2z2corres}. The free fermionic representation
is constructed at enhanced symmetry point in the toroidal moduli space, 
and deformation away from the free fermionic point are obtained by 
adding world--sheet Thirring interactions among the worldsheet 
fermions \cite{Thirring1987}. This construction produces a large space
of phenomenological three generation models 
\cite{fsu5, slm, PSmodels, newslm, 
lrs, psclass, su62, fsu5class, slmclass, lrsclass} 
that can be used to 
explore the physics of the Standard Model and its extensions.
Details of the free fermionic formulation and of the phenomenological 
three generation models are given in the references and will not
be repeated. Only the features relevant for the question of
sterile neutrinos and neutrino masses will be highlighted here.

In the free fermionic formulation of the light--cone heterotic--string in four 
dimensions there are 20 left--moving, and 44 right--moving, worldsheet
real fermions, 
The sixty--four free fermions are denoted by 
$$\{
\psi^{1,2}, 
(\chi, y, \omega)^{1,\cdots,6}\vert
({\bar y}, {\bar\omega} )^{1,\cdots,6},
{\bar\psi}^{1,\cdots,5}, 
{\bar\eta}^{1,2,3},
{\bar\phi}^{1,\cdots,8}
\},
$$
where 32 of the right--moving real fermions 
are grouped into 16 complex fermions that 
produce the Cartan generators of the rank 
16 gauge group. 
Here ${\bar\phi}^{1,\cdots,8}$ are the Cartan generators
of the rank eight hidden sector gauge group and ${\bar\psi}^{1,\cdots,5}$ are the 
Cartan generators of the $SO(10)$ GUT group. 
The complex worldsheet fermions 
${\bar\eta}^{1,2,3}$ generate three Abelian
currents, $U(1)_{1,2,3}$,
in the Cartan subalgebra of the observable gauge 
group with $U(1)_\zeta$ being their linear combination
\begin{equation}
U(1)_\zeta ~=~ U(1)_1+U(1)_2+U(1)_3~.
\label{uzeta}
\end{equation}
Models in the free fermionic formulation are constructed
by specifying a set of boundary condition basis vectors that specify the 
transformation properties of the 64 worldsheet fermions around the 
noncontractible loops of the worldsheet torus, 
$B=\{v_1, v_2, v_3, \cdots\}$, and 
the one--loop Generalised GSO (GGSO) phases in the partition function
$\cc{v_i}{v_j}$ \cite{fff}. 
The basis vectors spans an additive group $\Xi$, that contains
all possible linear combinations of the basis vectors
$\Xi= \sum_kn_kv_k$, where $n_k=0, \cdots, N_{v_k} -1$, and $N_{v_k}$ 
denote the order of each of the basis vectors. The physical 
representations in the Hilbert space of a given sector $\xi\in\Xi$ are 
obtained by acting on the vacuum with fermionic and bosonic 
oscillators and by applying the GGSO projections. 
The $U(1)$ charges of the physical states
with respect to the Cartan generators of the 
four dimensional gauge group are given by
$$Q(f) = \frac{1}{2}\xi(f) + F_\xi(f), $$
where $\xi(f)$ is the boundary condition of the complex 
worldsheet fermion $f$ in the sector $\xi$, and $F_\xi(f)$ is 
a fermion number operator \cite{fff}.
The phenomenological properties of the string model are 
extracted by calculating tree-level and higher order terms in the 
superpotential and by analysing its flat directions \cite{nrt}.

The free fermionic formulation of the heterotic--string produced a 
large space of phenomenological models that share the underlying 
$Z_2\times Z_2$ orbifold structure, with differing unbroken $SO(10)$ 
subgroups. The construction of the models can be viewed in two stages. 
The first consist of the basis vectors that preserve the $SO(10)$ symmetry. 
The models at this stage possess (2,0) worldsheet supersymmetry, with 
$N=1$ spacetime supersymmetry, and a number of vectorial and spinorial 
representations of $SO(10)$. The second stage of construction consist of 
the inclusion of the basis vectors that break the $SO(10)$ gauge group 
to a subgroup. The $SO(10)$ breaking vectors correspond to Wilson
line breaking in the corresponding $Z_2\times Z_2$ orbifold models. 
The $SO(10)$ preserving basis vectors are typically denoted by 
$\{b_i\}$ $i=1,2,3...$, whereas those that break the $SO(10)$ symmetry
are denoted by small Greek letters $\{\alpha, \beta, \gamma ...\}$.  
In all these models the unbroken $SO(10)$ subgroup contains an unbroken
combination of the $SO(10)$ Cartan generators, beyond the Standard Model
gauge group. This additional combination must therefore be broken
in the effective field theory low energy limit of the string models. 
The only available fields in the string models to achieve this 
breaking are fields that arise from the spinorial representation 
of $SO(10)$ and its conjugate. The case of the Standard--like 
models also contains exotic states that can be employed toward that
end. This distinction will not be important in the following. 
The weak hypercharge combination and the additional unbroken $U(1)$
are given by 
\beqn
U(1)_Y & =  & \frac{1}{3}U(1)_C ~+~ \frac{1}{2} U(1)_L,
\label{u1y}
\\
~~~ 
U(1)_{Z^\prime} & = & ~~U(1)_C ~-~  ~~U(1)_L ,
\label{u1zprime}       
\eeqn
where $U(1)_C$ and $U(1)_L$ are\footnote{$U(1)_C={3\over2}U(1)_{B-L};
U(1)_L=2U(1)_{T_{3_R}}.$} 
defined in terms of the worldsheet charges by
\beq
Q_C=Q({\bar\psi}^1)+Q({\bar\psi}^2)+ Q({\bar\psi}^3)~
{\rm and } ~
Q_L=Q({\bar\psi}^4)+Q({\bar\psi}^5).
\label{qcql}
\eeq
The string models contain additional unbroken $U(1)$ symmetries
and an unbroken hidden sector gauge group, the details of which are 
not crucial for the general discussion here. Aside from the linear 
combination given by $U(1)_\zeta$ in eq. (\ref{uzeta}), which arises 
from the breaking $E_6\rightarrow SO(10)\times U(1)_\zeta$. 
This breaking pattern is a generic consequence of the breaking 
of the worldsheet supersymmetry from $(2,2)$ to $(2,0)$ and 
is of vital importance in the ensuing discussion. The point is
that a generic consequence of this breaking is that $U(1)_\zeta$
is anomalous in most of the phenomenological free fermionic string vacua, 
and therefore cannot remain unbroken below the string scale. 
The sole example in which it is anomaly free is the string 
model of ref. \cite{frzprime}.

The entire spectrum of specific free fermionic string models is 
typically derived by applying the generalised GGSO projections. 
Here I focus on the generic features of the spectrum relevant for the 
question of sterile neutrinos. The three chiral generations are
obtained in these models from the three twisted sectors of the 
$Z_2\times Z_2$ orbifold, which are denoted in the free fermionic
models as $B_{1,2,3}$. The Standard Model electroweak Higgs doublet
may be obtained from the untwisted--sector or the twisted sectors.
In either case fermion mass terms are obtained from coupling of the 
chiral generations to the Higgs doublets and quasi-realistic mass spectrum
may be generated. Additionally, the models contain numerous Standard Model
and $SO(10)$ singlet states that may transform under the hidden gauge 
group and are charged under the additional $U(1)$ symmetries.
Typically, the models also contain a number of states that are 
singlets of the entire four dimensional gauge group. In some
models the number of such singlets is correlated with the
total number of generations and anti--generations.

In all the free fermionic models 
the breaking of the $U(1)$ symmetry in equation (\ref{u1zprime}) 
requires the existence of heavy Higgs fields ${\cal N}$
and ${\overline{\cal N}}$, that are obtained from the 
spinorial {\textbf 16} and $\overline{\textbf 16}$ representations of 
$SO(10)$. In models with an intermediate non--Abelian gauge symmetry, 
like the flipped $SU(5)$ (FSU5) \cite{fsu5, fsu5class}, the Pati--Salam (PS)
\cite{PSmodels, psclass}, the Left--Right Symmetric (LRS) 
\cite{lrs, lrsclass}, and the $SU(6)\times SU(2)$ (SU62) \cite{su62} 
models, the heavy Higgs fields break the non--Abelian gauge symmetry, 
whereas in the Standard--like Models (SLM) \cite{slm, slmclass}
they only break the extra $U(1)$ eq. (\ref{u1zprime}). 
In the case of the FSU5 and SU62 models the breaking is constrained 
to be of the order of the GUT scale, whereas in the other cases
it may be lower. The VEV of the ${\overline{\cal N}}$ field 
also generate the Majorana mass term of the right--handed neutrino 
in these models. 

The structure of the neutrino mass matrix is
quite generic in models inspired from the free fermionic models. 
The terms in the superpotential,
in term of component fields, 
that generate the neutrino mass matrix are (see {\it e.g.}
\cite{tauneutrinomass}),
\beq
L_iN_j{\bar h}~~~,~~~N_i{\overline {\cal N}}\phi_j~~~,~~~\phi_i\phi_j\phi_k~,
\label{supterms}
\eeq
where $N_i$, $L_i$ and $\phi_i$, with $i,j,k=1,2,3$ 
the right--handed neutrinos; 
are the chiral lepton doublets; 
and three $SO(10)$ singlet fields,
respectively;
${\bar h}$ is the electroweak Higgs doublet and ${\overline {\cal N}}$ is the
component of the heavy Higgs field that breaks $U(1)_{Z^\prime}$ in eq. (\ref{u1zprime}). 
All these states appear in the string models, possibly
as components of larger representation in as, {\it e.g.}, 
the PS and SU62 models. 
Generally, the neutrino seesaw matrix has the form 
\begin{equation}
{\left(
\begin{matrix}
                 {\nu_i}, &{N_i}, &{\phi_i}
\end{matrix}
   \right)}
  {\left(
\begin{matrix}
         0   &       (M_{_D})_{ij}    &             0                 \\
  (M_{_D})_{ij}&          0            & \langle{\overline {\cal N}}\rangle_{ij} \\
          0  &\langle{\overline{\cal N}}\rangle_{ij} & \langle\phi\rangle_{ij} \\
\end{matrix}
   \right)}
  {\left(
\begin{matrix}
                 {\nu_j}  \cr
                 {N_j}\cr
                 {\phi_j} \cr
\end{matrix}
   \right)},
\label{nmm}
\end{equation}
where $M_{_D}$ is the Dirac mass matrix arising from the first term in eq. 
(\ref{supterms}). The Dirac mass matrix is proportional to the up--quark mass
matrix due to the underlying $SO(10)$ symmetry  \cite{tauneutrinomass}. 
Taking the mass matrices 
to be diagonal the mass eigenstates are primarily $\nu_i$, $N_i$ and $\phi_i$ 
with negligible mixing and with the eigenvalues 
\beq
m_{\nu_j} \sim  
\left(
{{k M^j_u} \over {\langle{\overline{\cal N}}\rangle  }    }
\right)^2 \langle{\phi}\rangle~~~,~~~
\qquad m_{N_j},m_{\phi} \sim \langle{\overline {\cal N}}\rangle~.
\label{neutrinomasseigen}
\eeq
where $k$ is a renormalisation factor due to RGE evolution. 
The important point to consider
is whether any of the Standard Model singlet fields may serve the 
role of a sterile neutrino, 
{\it i.e.} it has to remain light and mix with the active neutrinos. 
The obvious observation
is that in general the answer is negative. While the string models 
contain numerous 
Standard Model and $SO(10)$ singlet fields, they appear mostly as 
vector--like fields 
and would therefore receive heavy mass along supersymmetric 
flat directions. From 
eq. (\ref{neutrinomasseigen}) it is seen that the mass eigenvalues correspond
to three light active neutrinos and six massive states with masses of the 
order of the seesaw scale. Therefore, there are no sterile neutrinos 
in these cases. 

The question of light neutrino masses was analysed in some 
detail in concrete string 
derived models in refs. \cite{AR1992, naheneutrinos}. Mass terms in the 
superpotential are obtained from renormalisable and nonrenormalisable terms 
by calculating correlators between vertex operators \cite{nrt}. 
The models typically contain an anomalous $U(1)$ gauge symmetry
that breaks supersymmetry at the string scale and destabilises the vacuum. 
The vacuum is stabilised by assigning Vacuum Expectation Values to Standard Model
singlet fields in the string massless spectrum, along 
supersymmetric $F$-- and $D$--flat
directions. Some of the nonrenormalisable operators then become renormalisable 
operators in the effective low energy field theory below the string scale. 
In this process many of the Standard Model singlets
receive heavy mass and decouple from the low energy spectrum. 
The seesaw mass matrix in eq. (\ref{nmm}) requires that some of the 
singlet fields obtain 
intermediate mass. As the singlet fields are typically vector--like 
this presents a major 
difficulty in generating the seesaw mechanism in the string models, let alone 
allowing for the existence of sterile neutrinos.
A priori it is not apparent that 
any of the non--chiral singlets can remain light.
I focus here on the analysis performed in ref, \cite{naheneutrinos} for the 
model in ref. \cite{newslm}. Details of the model and its spectrum 
are given in ref. 
\cite{newslm}. 

The set of fields in the model of ref. \cite{newslm}
that enter the seesaw mass matrix includes
the three right--handed neutrinos, $N_i$; 
the three left--handed neutrinos, $L_i$; 
and the set of Standard Model singlets. 
These include: $SO(10)$ singlets with $U(1)$ and hidden charges,
$\{\Phi_{45},\Phi_{1,2,3}^\pm,\Phi_{13},\Phi_{23},\Phi_{12},
T_i,V_i\}\oplus{\rm h.c.}$;
a set of $SU(3)\times SU(2)\times U(1)_Y$ singlets, $H_{13-14,17-20,23-26}$,
with $U(1)_{Z^\prime}$ charge;
three entirely neutral singlets $\xi_{1,2,3}$. 

In ref. \cite{naheneutrinos} a detailed study of the superpotential, 
with nonrenormalisable
terms up to order $N=10$, and of the supersymmetric flat directions 
was presented. 
The resulting neutrino mass matrix takes the approximate 
form \cite{naheneutrinos}

\begin{equation}
{\bordermatrix{
         & L_3& L_2& L_1& & N_3& N_2& N_1& & H_{23}&H_{25}&\Phi_{13}&\Phi_{45}&{\bar\Phi}_1^-&{\bar\Phi}_3^-&\Phi_2^+&{\bar\Phi}_2^+\cr
L_3      &  0 &  0 &  0 & & 0  &  0 &  r & &   0   &  0   &   0     &      0  &     0        &       0      &   0    &     0        \cr
L_2      &  0 &  0 &  0 & & r  &  0 &  r & &   0   &  0   &   0     &      0  &     0        &       0      &   0    &     0       \cr 
L_1      &  0 &  0 &  0 & & r  &  0 &  v & &   0   &  0   &   0     &      0  &     0        &       0      &   0    &     0       \cr
         &    &    &    & &    &    &    & &       &      &         &         &              &              &        &             \cr
N_3      &  0 & r  &  r & & 0  &  0 &  0 & &   0   &  x   &   0     &      0  &     0        &       0      &   0    &     0        \cr
N_2      &  0 & 0  &  0 & & 0  &  0 &  0 & &   z   &  0   &   u     &      z  &      u       &       u      &   z    &     u        \cr
N_1      &  r & r  &  v & & 0  &  0 &  0 & &   0   &  w   &   0     &      0  &     0        &       0      &    0   &     0       \cr
         &    &    &    & &    &    &    & &       &      &         &         &              &              &        &             \cr
H_{23}   &  0 &  0 &  0 & & 0  &  z &  0 & &   p   &  0   &   x     &      p  &     x        &       x      &    x   &      x       \cr
H_{25}   &  0 &  0 &  0 & & x  &  0 &  w & &   0   &  0   &   0     &      0  &     0        &       0      &    0   &      0       \cr
\Phi_{13}&  0 &  0 &  0 & & 0  &  u &  0 & &   x   &  0   &   q     &      x  &     y        &       y      &    0   &      y       \cr
\Phi_{45}&  0 &  0 &  0 & & 0  &  z &  0 & &   p   &  0   &   x     &      p  &     x        &       x      &    x   &      x       \cr
\Phi_1^-&  0  &  0 &  0 & & 0  &  u &  0 & &   x   &  0   &   y     &      x  &     q        &       q      &    q   &      q       \cr
\Phi_3^-&  0  &  0 &  0 & & 0  &  u &  0 & &   x   &  0   &   y     &      x  &     q        &       q      &    q   &      q       \cr
\Phi_2^+&  0  &  0 &  0 & & 0  &  z &  0 & &   x   &  0   &   0     &      x  &     q        &       q      &    0   &      x       \cr
{\bar\Phi}_2^+
        &  0  &  0 &  0 & & 0  &  0 &  0 & &   x   &  0   &   y     &      x  &     q        &       q      &    x   &      q     \cr}}
\nonumber\\
\nonumber\\
\label{stringderivedseesaw}
\end{equation}
where 
\beqn
&& r\sim{10^{-6}}{\rm GeV}~~~v\sim10^2{\rm GeV}~~~
w\sim10^{6}{\rm GeV}~~~q\sim10^{7}{\rm GeV}~~~u\sim10^{8}{\rm GeV}~\nonumber\\
&& x\sim10^{9}{\rm GeV}~~~z\sim10^{10}{\rm GeV}~~~p\sim10^{11}{\rm GeV}~~~y\sim10^{13}{\rm GeV}~
\nonumber
\eeqn
It is important to emphasise that this solution is not aimed to produce a realistic mass and 
mixing spectrum for the fermionic fields, but merely to explore its qualitative features. 
The mass matrix (\ref{stringderivedseesaw}) has the mass
eigenvalues $\{1.7\times10^{13},1.7\times10^{13},2\times10^{11},
1\times10^{10},9\times10^9,1\times10^9,1\times10^9,5\times10^6,
101,101,17.5,17.5,0.02,2.4^{-8}\}{\rm GeV}$. The lightest eigenvalue, of order $10{\rm eV}$,
is predominantly $L_3$. The lightest singlet states, with order $10\%$ mixing with $L_2$,
are two nearly degenerate combinations of $\sim70\%$ mixture of $\Phi_1^-$ and $\Phi_3^-$, 
and mass of order $17.5{\rm GeV}$. 
The remaining spectrum is readily analyzed and contains 
mixtures of the right-handed neutrinos and the $SO(10)$ singlets, 
with heavier masses. 
The main demonstration from the analysis
is the lesson that although a simple and elegant 
mechanism for the neutrino spectrum can be motivated 
from string theory in the form of (\ref{nmm}),
generating it from string models is not straightforward 
The main difficulty is to understand how the 
singlet masses can be protected from being too massive.
This problem already appears when trying to generate a seesaw mechanism for the 
three active neutrinos, let alone for any additional sterile neutrinos
at a comparable mass scale. 

Several comments are in order. 
The picture described above is generic in the string GUT models. 
The common features include the existence of an anomalous $U(1)$ symmetry
and proliferation of Standard Model singlets in the massless spectrum that 
nevertheless gain heavy or intermediate mass in the effective field theory
limit. 
Generation of quasi--realistic fermion mass matrices typically requires elaborate
solutions of the flat directions \cite{fhckm}. 
This in turn generate intermediate mass terms for the Standard Model singlet fields, 
and therefore eliminates them from being candidate sterile neutrino states. 
We may also consider the possibility of generating Majorana mass terms for the 
right--handed neutrinos from the nonrenormalisable terms
\beq
N_iN_j{\overline {\cal N}}{\overline {\cal N}},
\label{majornn}
\eeq
where ${\overline {\cal N}}$ obtains a VEV that breaks the $U(1)_{Z^\prime}$ 
in eq. (\ref{u1zprime}). 
In the SLM model of ref. \cite{newslm} this requires the utilization of
the exotic $H$ fields that carry $-1/2 Q(Z^\prime)$ with respect to the
charge of the right--handed neutrino, as the ${\overline {\cal N}}$ 
state is not obtained in the massless spectrum. 
The needed terms are of the form $NNHHHH\phi^n$. In the model of ref.
\cite{newslm} such terms were not found up to $N=14$
and hence cannot induce the seesaw mechanism. 
In the FSU5 \cite{fsu5, fsu5class}, PS \cite{PSmodels, psclass}, 
LRS \cite{lrs, lrsclass} and SU62 \cite{su62} models 
the ${\overline {\cal N}}$ states do arise in the spectrum and may generate
terms of the form of eq. (\ref{majornn}). However, while this 
may contribute to the implementation of the seesaw mechanism, 
its role in allowing for light sterile neutrinos in the string derived models
is yet to be examined. 

From the discussion above it is clear that existence of sterile neutrinos 
at a mass scale of the order of the active neutrinos and with substantial mixing
with them, is rather problematic in string derived GUT models. This situation 
is, however, not without hope.
A model that may present an alternative scenario 
is the string derived $Z^\prime$ model of ref. 
\cite{frzprime}, where the unbroken $SO(10)$ subgroup is $SO(6)\times SO(4)$. 
The model is self dual under the spinor-vector duality of ref. 
\cite{spinvecduality}. 
As a result the chiral spectrum in this models forms complete $E_6$ 
representation, 
and consequently the $U(1)_\zeta$ combination of eq. (\ref{uzeta}) is 
anomaly free and may remain unbroken below the string scale. 
The gauge symmetry, however, is not enhanced, and spacetime vector
bosons arise in this model solely from the untwisted sector. 
The entire massless spectrum of the model, 
and the charges under the gauge group, 
are given in ref \cite{frzprime}. 
Tables \ref{tableb} and \ref{tablehi} provide a glossary of the 
states in the model and their charges under the 
$SU(4)\times SO(4)\times U(1)_\zeta$ gauge group.
We note in particular the existence of the seven
$S$ and four $\overline{S}$ states that are singlets 
of $SO(10)$ and are charged under $U(1)_\zeta$. There are therefore 
three chiral states that are singlets of $SO(10)$ and 
charged under $U(1)_\zeta$. They are left--over components from 
the {\textbf 27} and $\overline{\textbf27}$ representations of $E_6$. 

\begin{table}[!h]
\begin{tabular}{|c|c|c|c|}
\hline
Symbol& Fields in \cite{frzprime} & 
                         $SU(4)\times{SU(2)}_L\times{SU(2)}_R$&${U(1)}_{\zeta}$\\
\hline
$\FL$ &        $F_{1L},F_{2L},F_{3L}$&$\left({\bf4},{\bf2},{\bf1}\right)$&$+\oh$\\
$\FR$ &$F_{1R}$&$\left({\bf4},{\bf1},{\bf2}\right)$&$-\oh$\\
$\FbR$&${\bar F}_{1R},{\bar F}_{2R},{\bar F}_{3R},{\bar F}_{4R}$
                             &$\left({\bf\bar 4},{\bf1},{\bf2}\right)$&$+\oh$\\
$h$   &$h_1,h_2,h_3$&$\left({\bf1},{\bf2},{\bf2}\right)$&$-1$\\
$\Delta$&$ D_1,\dots,  D_7$&$\left({\bf6},{\bf1},{\bf1}\right)$&$-1$\\
$\bar\Delta$&$\Db_1,\Db_2,\Db_3,\Db_6$&$\left({\bf6},{\bf1},{\bf1}\right)$&$+1$\\
$S$&$\Phi_{12},\Phi_{13},\Phi_{23},\chi^+_1,\chi^+_2,\chi^+_3,\chi^+_5$
&$\left({\bf1},{\bf1},{\bf1}\right)$&$+2$\\
$\Sb$&$\bar\Phi_{12},\bar\Phi_{13},\bar\Phi_{23},\bar\chi^+_4$&$\left({\bf1},{\bf1},{\bf1}\right)$&$-2$\\
$\phi$&$\phi_1,\phi_2$&$\left({\bf1},{\bf1},{\bf1}\right)$&$+1$\\
$\phib$&$\bar\phi_1,\bar\phi_2$&$\left({\bf1},{\bf1},{\bf1}\right)$&$-1$\\
$\zeta$&$\Phi_{12}^-,\Phi_{13}^-,\Phi_{23}^-,\bar\Phi_{12}^-,\bar\Phi_{13}^-,\bar\Phi_{23}^-$&$\left({\bf1},{\bf1},{\bf1}\right)$&$\hphantom{+}0$\\
&$\chi_1^-,\chi_2^-,\chi_3^-,\bar\chi_4^-,\chi_5^-$&$$&$$\\
&$\zeta_i,\bar\zeta_i,i=1,\dots,9$&$$&$$\\
&$\Phi_i,i=1,\dots,6$&$$&$$\\
\hline
\end{tabular}
\caption{\label{tableb}
Observable sector field notation and associated states in \cite{frzprime}.}
\end{table}
\begin{table}[!h]
\begin{center}
\begin{tabular}{|c|c|c|c|}
\hline
Symbol& Fields in \cite{frzprime} & ${SU(2)}^4\times SO(8)$&${U(1)}_{\zeta}$\\
\hline
$H^+$&$H_{12}^3$&$\left({\bf2},{\bf2},{\bf1},{\bf1},{\bf1}\right)$&$+1$\\
&$H_{34}^2$&$\left({\bf1},{\bf1},{\bf2},{\bf2},{\bf1}\right)$&$+1$\\
$H^-$&$H_{12}^2$&$\left({\bf2},{\bf2},{\bf1},{\bf1},{\bf1}\right)$&$-1$\\
&$H_{34}^3$&$\left({\bf1},{\bf1},{\bf2},{\bf2},{\bf1}\right)$&$-1$\\
$H$&$H_{12}^1$&$\left({\bf2},{\bf2},{\bf1},{\bf1},{\bf1}\right)$&$0$\\
&$H_{13}^i,i=1,2,3$&$\left({\bf2},{\bf1},{\bf2},{\bf1},{\bf1}\right)$&$0$\\
&$H_{14}^i,i=1,2,3$&$\left({\bf2},{\bf1},{\bf1},{\bf2},{\bf1}\right)$&$0$\\
&$H_{23}^1$&$\left({\bf1},{\bf2},{\bf2},{\bf1},{\bf1}\right)$&$0$\\
&$H_{24}^1$&$\left({\bf1},{\bf2},{\bf1},{\bf2},{\bf1}\right)$&$0$\\
&$H_{34}^i,i=1,4,5$&$\left({\bf1},{\bf1},{\bf2},{\bf2},{\bf1}\right)$&$0$\\
$Z$&$Z_i,i=1,\dots,$&$\left({\bf1},{\bf1},{\bf8}\right)$&$0$\\
\hline
\end{tabular}
\end{center}
\caption{\label{tablehi}
Hidden sector field notation and associated states in \cite{frzprime}. }
\end{table}
The string model is obtained by trawling a self--dual model 
under the spinor--vector duality at the $SO(10)$ level, 
{\it i.e.} prior to breaking the $SO(10)$ symmetry to the Pati--Salam 
subgroup. 
Self--duality under the spinor--vector duality plays a key in
the construction of the string model with anomaly free $U(1)_\zeta$, 
together with $E_6$ embedding of the $U(1)_\zeta$ charges. In this respect 
it is noted that $U(1)_\zeta$ is anomaly free 
in the LRS heterotic--string models as well \cite{lrs, lrsclass}. However, 
in the LRS models the $U(1)_\zeta$ charges of the Standard Model states 
do not possess the $E_6$ embedding. The VEV of ${\cal N}$ and 
$\overline{\cal N}$ 
leaves unbroken the same $U(1)$ combination of $U(1)_C$, $U(1)_L$ and 
$U(1)_\zeta$, 
but in the case of the LRS models its breaking at low scales would lead to 
contradiction with $\sin^2\theta_W(M_Z)$ and $\alpha_s(M_Z)$ \cite{fmgcu}.
An alternative to the construction of ref \cite{frzprime} is the 
construction proposed 
in ref. \cite{afm} that essentially amounts to an alternative 
$E_6$ breaking pattern, 
as follows. The contributions to the observable gauge symmetry in the 
free fermionic models may come from two sectors. The first is the untwisted 
sector and the second is the $x$--sector that enhances the ten dimensional 
observable $SO(16)$ to $E_8$. In most of the phenomenological free fermionic
models the spacetime vector bosons from the $x$--sector are projected out, 
which results in the symmetry breaking pattern 
$E_6\rightarrow SO(10)\times U(1)_\zeta$, 
and $U(1)_\zeta$ becoming anomalous. In ref. \cite{afm} an alternative symmetry
breaking pattern is proposed that retains the spacetime vector bosons from the 
$x$--sector and may allow for $U(1)_\zeta$ to remain anomaly free. 
An explicit string
derived model that realises this symmetry breaking pattern is 
the three generation
$SU(6)\times  SU(2)$ GUT model of ref. \cite{su62}. In this model 
$U(1)_\zeta$ is anomaly free by virtue of its embedding in the GUT group. 
It should be remarked that the precise combination of $U(1)_{1,2,3}$ that 
forms $U(1)_\zeta$ may differ from the one in eq. (\ref{uzeta}), up to signs.
This is the case, for example, in the model of ref. \cite{su62}. 
The important properties of $U(1)_\zeta$ are that: 
it is the family universal combination; it is anomaly free;
and the charges of the Standard Model states admit the $E_6$ embedding.
The models of refs. \cite{frzprime}
and \cite{su62} demonstrate how this is achieved, either by exploiting the 
self--duality under the spinor--vector duality, as in the model ref. 
\cite{frzprime}, or embedding $U(1)_\zeta$ in a non--Abelian group, 
as in the model of ref. \cite{su62}.

In either of these cases the massless string spectrum contains the 
fields required to break the GUT symmetry to the Standard Model.
I focus here on the model of ref. \cite{frzprime}. 
The observable and hidden gauge groups at the 
string scale are produced by untwisted sector states and are given by: 
\beqn
{\rm observable} ~: &~~~~~~~~SO(6)\times SO(4) \times 
U(1)_1 \times U(1)_2\times U(1)_3 \nonumber\\
{\rm hidden}     ~: &SO(4)^2\times SO(8)~~~~~~~~~~~~~~~~~~~~~~~~\nonumber
\eeqn
Additional spacetime vector bosons that may enhance 
the observable and hidden gauge symmetries are projected 
out in this model due to the choice of GGSO projection coefficients.
There are two anomalous 
$U(1)$s in the string model with 
\beq
{\rm Tr}U(1)_1= 36 ~~~~~~~{\rm and}~~~~~~~{\rm Tr}U(1)_3= -36. 
\label{u1u3}
\eeq
Therefore, the $E_6$ combination in eq. (\ref{uzeta}) is anomaly free
and can be as a component of an extra $Z^\prime$ below the string scale.
As noted from table \ref{tablehi} the model contains hidden
sector vector--like states, that include: 
four bidoublets denoted by $H^\pm$ with $Q_{\zeta}=\pm1$ charges;
12 neutral bidoublets denoted by $H$; 
and five states that transform in the $8$ representation of the 
hidden $SO(8)$ gauge group with $Q_\zeta=0$.
The observable $SO(6)\times SO(4)$ gauge symmetry in the model is
broken by the VEVs of the heavy Higgs fields ${\cal H}=F_R$ and 
$\overline{\cal H}$, which is a linear combination of the four 
$\bar{F}_R$ fields. The decomposition of these fields in terms of the 
Standard Model gauge group factors is given by: 
\begin{align}
\overline{\cal H}({\bf\bar4},{\bf1},{\bf2})& \rightarrow u^c_H\left({\bf\bar3},
{\bf1},\frac 23\right)+d^c_H\left({\bf\bar 3},{\bf1},-\frac 13\right)+
                            {\overline {\cal N}}\left({\bf1},{\bf1},0\right)+
                             e^c_H\left({\bf1},{\bf1},-1\right)
                             \nonumber \\
{\cal H}\left({\bf4},{\bf1},{\bf2}\right) & 
\rightarrow  u_H\left({\bf3},{\bf1},-\frac 23\right)+
d_H\left({\bf3},{\bf1},\frac 13\right)+
              {\cal N}\left({\bf1},{\bf1},0\right)+ 
e_H\left({\bf1},{\bf1},1\right)\nonumber
\end{align}
The VEVs along the ${\cal N}$ and $\overline{\cal N}$ directions leave the 
unbroken combination
\beq
U(1)_{{Z}^\prime} ~=~
{1\over {5}} (U(1)_C - U(1)_L) - U(1)_\zeta
~\notin~ SO(10),
\label{uzpwuzeta}
\eeq
that may remain unbroken below the string scale provided that $U(1)_\zeta$ is
anomaly free. The cancellation of the $U(1)_{Z^\prime}$ anomalies requires 
the presence of the vector--like quarks $\{D_i, \overline{D}_i\}$ and leptons
$\{H_i, {\bar H}_i\}$, that arise 
from the vectorial $10$ representation of $SO(10)$, as well as the $SO(10)$ 
singlets $S_i$ in the $27$ of $E_6$. 
The spectrum below the Pati--Salam breaking
scale is displayed schematically in table \ref{table27rot}. 
The three right--handed 
neutrino $N_L^i$ states become massive at the $SU(2)_R$ breaking scale, 
which generates the seesaw mechanism via either eq. (\ref{nmm})
or via the nonrenormalisable term in eq. (\ref{majornn}). 
I assume here that the spectrum is supersymmetric
below the $SU(2)_R$ breaking scale, and allow  
for the possibility that the spectrum contains an additional
pair of vector--like electroweak Higgs doublets. 
Additionally, the existence of light states $\zeta_i$, 
that are neutral under the 
$SU(3)_C\times SU(2)_L\times U(1)_Y\times U(1)_{Z^\prime}$ low
scale gauge group, is allowed. The states $\phi$ and ${\bar\phi}$ are 
exotic Wilsonian states \cite{ews}, that match the 
${\phi_{1,2}}$ and ${\bar\phi_{1,2}}$ in table \ref{tableb}. 
The $U(1)_{Z^\prime}$ 
gauge symmetry can be broken at low scales by the VEV of the $SO(10)$ singlets
$S_i$ and/or ${\phi_{1,2}}$. 

\begin{table}[!h]
\noindent 
{\small
\begin{center}
{\tabulinesep=1.2mm
\begin{tabu}{|l|cc|c|c|c|}
\hline
Field &$\hphantom{\times}SU(3)_C$&$\times SU(2)_L $
&${U(1)}_{Y}$&${U(1)}_{Z^\prime}$  \\
\hline
$Q_L^i$&    $3$       &  $2$ &  $+\frac{1}{6}$   & $-\frac{2}{5}$   ~~  \\
$u_L^i$&    ${\bar3}$ &  $1$ &  $-\frac{2}{3}$   & $-\frac{2}{5}$   ~~  \\
$d_L^i$&    ${\bar3}$ &  $1$ &  $+\frac{1}{3}$   & $-\frac{4}{5}$  ~~  \\
$e_L^i$&    $1$       &  $1$ &  $+1          $   & $-\frac{2}{5}$  ~~  \\
$L_L^i$&    $1$       &  $2$ &  $-\frac{1}{2}$   & $-\frac{4}{5}$  ~~  \\
%
\hline
$D^i$       & $3$     & $1$ & $-\frac{1}{3}$     & $+\frac{4}{5}$  ~~    \\
${\bar D}^i$& ${\bar3}$ & $1$ &  $+\frac{1}{3}$  &   $+\frac{6}{5}$  ~~    \\
$H^i$       & $1$       & $2$ &  $-\frac{1}{2}$   &  $+\frac{6}{5}$ ~~    \\
${\bar H}^i$& $1$       & $2$ &  $+\frac{1}{2}$   &   $+\frac{4}{5}$   ~~  \\
\hline
$S^i$       & $1$       & $1$ &  ~~$0$  &  $-2$       ~~   \\
\hline
$h$         & $1$       & $2$ &  $-\frac{1}{2}$  &  $-\frac{4}{5}$  ~~    \\
${\bar h}$  & $1$       & $2$ &  $+\frac{1}{2}$  &  $+\frac{4}{5}$  ~~    \\
%
%
%
\hline
$\phi$       & $1$       & $1$ &  ~~$0$         & $-1$     ~~   \\
$\bar\phi$       & $1$       & $1$ &  ~~$0$     & $+1$     ~~   \\
\hline
\hline
$\zeta^i$       & $1$       & $1$ &  ~~$0$  &  ~~$0$       ~~   \\
\hline
\end{tabu}}
\end{center}
}
\caption{\label{table27rot}
\it
Spectrum and
$SU(3)_C\times SU(2)_L\times U(1)_{Y}\times U(1)_{{Z}^\prime}$ 
quantum numbers, with $i=1,2,3$ for the three light 
generations. The charges are displayed in the 
normalisation used in free fermionic 
heterotic--string models. }
\end{table}

The extended set of fields appearing in the seesaw mass matrix in this model
are $\{L_i, N_i, H_i, S_i, h, {\bar h} , \phi, {\bar\phi}, \zeta_i\}$, and 
we wish to examine scenarios leading to a sterile neutrino 
with substantial mixing with the active neutrino. 
The allowed renormalisable couplings among these fields, 
contributing to the seesaw mass matrix,
in the model are 
\beq
L_iN_j{\bar h}, L_iN_j{\bar H}_k,~
N_i\zeta_j\overline{\cal N}, ~
H_i{\bar H}_jS_k, H_i{\bar h}S_k, ~
{\bar H}_ih\zeta_j,~
h{\bar h}\zeta_i,~
\phi{\bar\phi}\zeta_i,~
{\bar\phi}{\bar\phi}S_i,~
\zeta_i\zeta_j\zeta_k.
\label{couplings}
\eeq
as well as the nonrenormalisable coupling in eq. (\ref{majornn}). 
Restricting for simplicity solely to the set of chiral fields under 
the $U(1)_{Z^\prime}$ gauge symmetry generates the seesaw mass matrix
in eq. (\ref{sterilematrix}), per generation, 
\begin{equation}
{\bordermatrix{
               & L_i       & S_i           &  H_i        & \overline{H}_i   & N_i   \cr
L_i            &  0        &       0       &  0          & \lambda n       & \lambda v    \cr
S_i            &  0        &       0       & \lambda v_2 & \lambda v_3      & 0     \cr 
H_i            &  0        &  \lambda v_2  &  0          & z^\prime          & 0     \cr
\overline{H}_i &  \lambda n&  \lambda v_3  & z^\prime      & 0               & 0     \cr
N_i            &  \lambda v&       0       &  0          & 0               & \frac{~\overline{{\cal N}}^2}{M} 
\cr}},
\nonumber\\
\nonumber\\
\label{sterilematrix}
\end{equation}
which is
produced by the relevant couplings in eqs. (\ref{couplings}, \ref{majornn}). 
The analysis here is for only for illustration and performed for a single generation. 
The seesaw mass matrix in eq. (\ref{sterilematrix}) contains VEVs that break
three distinct symmetries: 
the VEVs that break the $SU(2)_R$ symmetry, 
$\overline{\cal N}$ and $n$;
the VEVs that break the $U(1)_{Z^\prime}$ symmetry denoted by 
$z^\prime$ in (\ref{sterilematrix}); 
and the VEVs that break the electroweak symmetry denoted by 
$v$, $v_2$ and $v_3$ in eq. (\ref{sterilematrix}).
Here $v$ is the VEV that produces the Dirac mass terms 
that couples between the left-- and 
right--handed neutrinos, and $v_2$ and $v_3$ are VEVs
along the lepton doublets $H$ and ${\bar H}$. 
The notation in eq. (\ref{sterilematrix}) expresses generically
the dependence of some mass entries on unknown Yukawa couplings. 
The aim here is not a detailed numerical analysis, but merely to 
provide an example that demonstrates the possibility of sterile 
neutrinos in string GUT construction. 
The mass matrix of the light eigenvalues can be obtained by 
defining the seesaw mass matrix eq. (\ref{sterilematrix}) in the form 
\begin{equation}\label{svss}
{\bf M}=\left(\begin{array}{cc}0&{\bf B}\\{\bf B}^T&{\bf J}\end{array}\right), 
\end{equation}
with 
\beqn
{\bf J} &=& 
\left(\begin{matrix}
  0          & z^\prime  & 0     \cr
 z^\prime    &    0      & 0     \cr
  0          &    0      & \frac{~\overline{{\cal N}}^2}{M} \cr
\end{matrix}\right),
\nonumber\\
& & \nonumber\\
{\bf B} &=&
\left(\begin{matrix}
 0          & \lambda n       & \lambda v    \cr
\lambda v_2 & \lambda v_3      & 0     \cr
\end{matrix}\right),
\nonumber 
\eeqn
where $M \approx 10^{18}GeV$ is related to the heterotic--string scale. 
The light eigenvalues and eigenvectors are approximately given by those of the 
matrix 
\beq
{\bf B} {\bf J}^{-1}{\bf B}^T \approx 
\left( 
\begin{matrix}
 \frac{M (\lambda v)^2}{\overline{\cal N}^2} &  & 
                        \frac{(\lambda n) (\lambda v_2)}{z^\prime}    \cr
                                           & & \cr
\frac{(\lambda n) (\lambda v_2)}{z^\prime} & &
                        \frac{2 (\lambda v_2)( \lambda v_3)}{z^\prime}    \cr
\end{matrix}
\right),
\eeq
where the $\lambda$ coupling appearing is schematic. 
Taking the parameters in eq. (\ref{sterilematrix}) to be given by: 
\beqn
\lambda v   &=& 1{\rm GeV};~\\
\lambda v_2 &=& 5\times 10^{-4}{\rm GeV}\approx m_e; ~\nonumber\\
\lambda v_3 &=& 5\times 10^{-4}{\rm GeV}; \approx m_e;~ \nonumber\\
\lambda n   &=& 5\times 10^{-4}{\rm GeV}; \approx m_e;~\nonumber\\
z^\prime    &=& 5\times 10^4{\rm GeV} = 50{\rm TeV};~\nonumber\\
\overline{\cal N} &=& 5\times10^{14}{\rm GeV},~\nonumber
\eeqn
produces two light eigenvalues with $m_1=10^{-2}$eV and 
$m_2=10^{-3}$eV and three massive states
with $m_3=m_4=50$TeV and $m_5=2.5\times 10^{11}$GeV. 
The heavy eigenstate is the right--handed 
neutrino. The two intermediate states are equal mixtures of the 
$H_i$ and ${\bar H}_i$ and 
the two light eigenstates are mixtures of the 
left--handed neutrino and $S_i$, 
with mixing angle $\sin\theta\approx 0.98$. 
This is of course only an illustrative scenario and
many other possibilities exist. However, it does 
demonstrate the possibility of producing 
sterile neutrinos in heterotic--string constructions. 
And more importantly, it suggests that the
existence of a sterile neutrino in this constructions may be 
correlated with a new Abelian gauge 
symmetry not far removed from scales currently probed in collider experiments. 

\section{Conclusions}

The results of the MiniBoonNE experiment provide evidence for the existence of 
sterile neutrinos in nature with substantial mixing with the active neutrinos. 
These results have profound implications for attempts to unify the Standard 
Model
with gravity. String theory provides a concrete framework where these 
implications 
can be studied.

In this paper I argued that sterile neutrinos arise naturally in large 
volume scenarios
and by implication in string constructions in which the gravitational 
scale may be as low as
the TeV scale. Of course, the precise nature of the sterile neutrinos 
in these constructions 
needs to be scrutinised, but it is clear that the sterile neutrinos in these models 
can affect various experimental observables that are being explored in contemporary experiments, 
{\it e.g.} lepton universality \cite{lvneutrinos}. 
I further argued that obtaining sterile neutrinos in string GUT constructions proves more
of a challenge, and investigated this question in perturbative heterotic--string 
models in the free fermionic formulation. I argued that generic phenomenological 
models in this category would not lead to sterile neutrinos at a mass scale 
comparable to the active neutrinos and with substantial mixing with them. 
The main reason being that the Standard Model singlet states in these models arise
in vector--like representation with respect to the unbroken four dimensional 
gauge group at the string scale. A way to guarantee that some singlets remain light and 
can act as sterile neutrinos is if they are chiral with respect to an additional
Abelian gauge symmetry, which is broken at low or intermediate scales. 
The construction of string models that allow for the required extra Abelian symmetry is, 
however, highly non--trivial and I discussed how it is realised in some free fermionic 
heterotic--string models. The results of the MiniBooNE experiment and its successors 
may therefore provide vital guidance in the quest for unification of gravity and the 
gauge interactions.

\section*{Acknowledgments}

I would like to thank the Simons Center for Geometry and Physics at State University of New York 
in Stony Brook for hospitality while part of this work was conducted.

\end{document}